\documentclass[a4paper,twocolumn]{revtex4}

\usepackage{amsmath}
\usepackage{amsfonts}
\usepackage{graphics}
\usepackage{enumerate}
\usepackage{graphicx}
\begin{document}

\title{Iterated dynamical maps in an ion trap.}
%\date{}
\author{M. Duncan$^1$,  J. Links$^1$, and G. J. Milburn$^2$.\\ \hspace{1cm}}
\address{$^1$School of Physical Sciences,\\
$^2$The Centre for Quantum Computer Technology, \\ The University of Queensland,QLD 4072 Australia.}
\begin{abstract}
Iterated dynamical maps offer an ideal setting to investigate quantum dynamical bifurcations and are well adapted to few-qubit quantum computer realisations.  We show that a single trapped ion, subject to periodic impulsive forces, exhibits  a rich structure of dynamical bifurcations derived from the Jahn-Teller Hamiltonian flow model.  We show that the entanglement between the oscillator and electronic degrees of freedom reflects the underlying dynamical bifurcation in a Floquet eigenstate. 
\end{abstract}
\maketitle

%\section{Introduction}
Iterated area-preserving maps are better adapted to physical implementations of quantum information processing than Hamiltonian flows. In fact one of the first quantum algorithms, the Grover search algorithm\cite{Grover}, may be regarded as a quantum description of an area-preserving map. Typically, a quantum algorithm consists of a sequence of elementary unitary operations on one or more two-level systems.  In the Grover search algorithm a simple product of unitary operators is iterated. While it is possible to simulate an arbitrary Hamiltonian flow as a quantum  circuit,  area-preserving maps are simulated more directly as an iterated sequence of unitary gates. Originally introduced by Poincar\'{e}, iterated maps have formed the core of studies in quantum chaos for many decades\cite{haake}. The model reported here is a simple example of a non trivial area-preserving map, based on the Jahn-Teller model\cite{jt} and is naturally adapted to an ion trap implementation of quantum information processing. 

The Jahn-Teller model describes a class of systems in which one or more particle coordinates 
are coupled to a two level system\cite{Eng72}. The $E\otimes\epsilon$ Jahn-Teller model is a minimal description in which two
harmonic oscillator coordinates are coupled to a two level system. We may write the Jahn -Teller $E\otimes\epsilon$ Hamiltonian as
\begin{equation}
H_0 ={\widetilde{\Delta}} s_z+ \frac{1}{2m}(p^2_x +p^2_y) + \frac{m \widetilde{\omega}^2}{2} (q^2_x +q^2_y)+\lambda_xq_x s_x+\lambda_y q_y s_y
\label{hamiltonian} 
\end{equation}
where $\{q_x,q_y,p_x,p_y\}$ are the conjugate position and momentum operators in the harmonic oscillator space, and $\{s_x,s_y,s_z\}$ are $su(2)$ operators acting on the states of the two-level internal degree of freedom.

The classical description of this model is a Hamiltonian flow in the phase-space of the  Cartesian product of the phase plane of the oscillators $\mathbb{R}^2$ and the spherical phase space of the two level system $S^2$, known as the Bloch sphere. The quantum model exhibits the {\em conical intersection} that has assumed importance in various biophysical models of vision\cite{CI}.  It is well known that the $E\otimes\epsilon$ Jahn-Teller model displays a classical bifurcation of fixed points as the coupling strength is varied and that the quantum analog displays a maximum 
in the entanglement at the same value of the coupling strength\cite{LevMut01,hines}  as the classical fixed point bifurcation. At this point there is a morphological change in the nature of the ground state. This is the few body analogue of a quantum phase transition in a true many body system.    

We do not consider the Hamiltonian flow model, but rather a related model, which 
we will call a {\em kicked} $E\otimes\epsilon$ model that is better adapted to an ion trap realisation. Instead of a time-independent Hamiltonian, 
we will consider a strongly time dependent version of the Hamiltonian. We 
assume that
 $$\lambda_{x,y}\mapsto \lambda_{x,y}\sum_n\delta(t-{n \tau_{x,y}})$$ A hamiltonian which is periodically varying in time is most naturally described in terms of the Floquet operator, a unitary operator that maps the dynamics over one period. 

 Two kicking interactions are periodically applied instantaneously and consecutively, after which there is a period of evolution $\tau$ under the  Hamiltonian flow associated with the confining potential. 
 
% 
% Given a time-dependent Hamiltonian of the general form $H=H_0+\widetilde{\lambda}(t)H_{1}$, states evolve under the unitary transformation (with $\hbar=1$)
%\begin{equation*}
%U(t)=\exp\left[-{i}\int_0^t H(s)ds\right].
%\end{equation*}
%For periodically kicked systems, where the time-dependent coupling is taken to be  
%$\widetilde{\lambda}=\lambda\sum_n\delta(t-n\tau)$, it is convenient to take a stroboscopic view of the dynamics through a discrete time map over the period $\tau$ of the interaction.  This linear map, referred to as the Floquet operator, takes the form
%\begin{equation*} U=\exp\left({-i} H_0\tau\right)\exp\left({-i}\lambda H_1\right)\end{equation*}
%For the present model with two kicked interactions we obtain the Floquet operator (\ref{floquet}). 
% 
 
 The unitary operator which discretely maps states over each period is the Floquet operator (setting $\hbar=1$)
\begin{align}
U&=\exp(-iH_0\tau)\exp(-i\lambda H_{x})\exp(-i\lambda H_{y}),
\nonumber\\
H_0 &={\widetilde{\Delta}} s_z+ \frac{1}{2m}(p^2_x +p^2_y) + \frac{m \widetilde{\omega}^2}{2} (q^2_x +q^2_y),\nonumber\\
%H_{x}&={\lambda}q_x s_x,\nonumber\\
H_{\chi}&=q_\chi s_\chi,\qquad \chi=x,\,y
\label{floquet} 
\end{align}
The unitary map (\ref{floquet}) is the kicked equivalent of the $E \otimes \epsilon$ Jahn--Teller model. We will fix the frequency of the harmonic potential, $\widetilde{\omega}$, the internal splitting energy, $\widetilde{\Delta}$, the periodicity of the kick $\tau$ and adopt units such that $m=\widetilde{\omega}^{-1}$. This leaves the kick coupling $\lambda$ as the tunable parameter. 

For each system observable $A$ evolving under a Hamiltonian $H$ the time evolution is
${d\langle A\rangle}/{dt}=i\langle[H,A]\rangle$.
From this expression, we can derive operator differential equations for the evolution of all seven operators under each of $H_0,\,H_{x},\,H_{y}$. According to Ehrenfest's theorem \cite{Sakurai}, in the classical limit expectation values of operators approach classical variables. Hence the differential operator equations become a set of classical equations in this limit.  Integrating these equations over one period $\tau$ gives three maps which we then compose into a single area-preserving map on the classical phase space
$M(\mathbb{R}^4\times \mathcal{S}^2)\mapsto \mathbb{R}^4\times\mathcal{S}^2$
where $\mathcal{S}^2$ denotes the Bloch sphere. 

The operator time evolution equations for $q_x,\,q_y,\,p_x,\,p_y,\,s_x,\,s_y,\,s_z$ can be determined under each of $\exp(-iH_0)$, $\exp(-iH_x)$, $\exp(-iH_y)$. Taking these to be classical differential equations, in each case we can integrate these to obtain three maps for the corresponding classical variables. Composing these three maps yields   
\begin{widetext}
\small
\begin{align}
q_x &\mapsto \bigl(p_x-\lambda\{s_x \cos(\lambda q_y)+s_z \sin(\lambda q_y)\}\bigr)\sin\omega+q_x\cos\omega, 
\qquad q_y \mapsto \bigl(p_y-\lambda s_y\bigr)\sin\omega+q_y\cos\omega, \label{map1} \\
p_x &\mapsto \bigl(p_x-\lambda\{s_x \cos(\lambda q_y)+s_z \sin(\lambda q_y)\}\bigr)\cos\omega-q_x\sin\omega,  
\qquad
p_y \mapsto \bigl(p_y-\lambda s_y\bigr)\cos\omega-q_y\sin\omega\\
s_x &\mapsto \bigl(\cos\Delta\cos(q_y\lambda)-\sin\Delta\sin(q_x\lambda)\sin(q_y\lambda)\bigr)s_x-\sin\Delta\cos(q_x\lambda)s_y\nonumber\\
&\mbox{} \ \ \ \ \ \ \ \ \ \ +\bigl(\cos\Delta\sin(q_y\lambda)+\sin\Delta\sin(q_x\lambda)\cos(q_y\lambda)\bigr)s_z\\
s_y &\mapsto \bigl(\cos\Delta\sin(q_x\lambda)\sin(q_y\lambda)+ \sin\Delta\cos(q_y\lambda)\bigr)s_x\nonumber\\
 &\mbox{} \ \ \ \ \ \ \ \ \ \ +\cos\Delta\cos(q_x\lambda)s_y+\bigl(-\cos\Delta\sin(q_x\lambda_a)\cos(q_y\lambda)+\sin\Delta\sin(q_y\lambda)\bigr)s_z\\
s_z &\mapsto -\sin(q_y\lambda)\cos(q_x\lambda)s_x+\sin(q_x\lambda)s_y+\cos(q_y
\lambda)\cos(q_x\lambda)s_z
\label{map2} 
\end{align}
\normalsize
\end{widetext}
%It is readily verified that (\ref{fp}) is a fixed point of the above map, and more generally all fixed points are constrained to lie on the submanifold $p_\chi=-\tan(\omega/2) q_\chi$ for $\chi=x,\,y$. Expanding the map about (\ref{fp}) to quadratic order, it is found that the stability of the fixed point changes when (\ref{bif}) holds. 

\begin{figure}
\begin{tabular}{cc}
\includegraphics[scale=0.55]{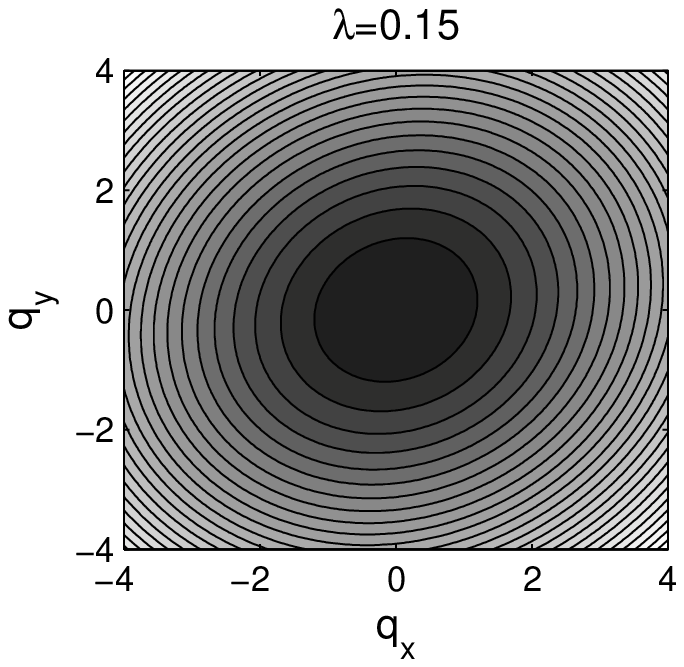} & 
\includegraphics[scale=0.55]{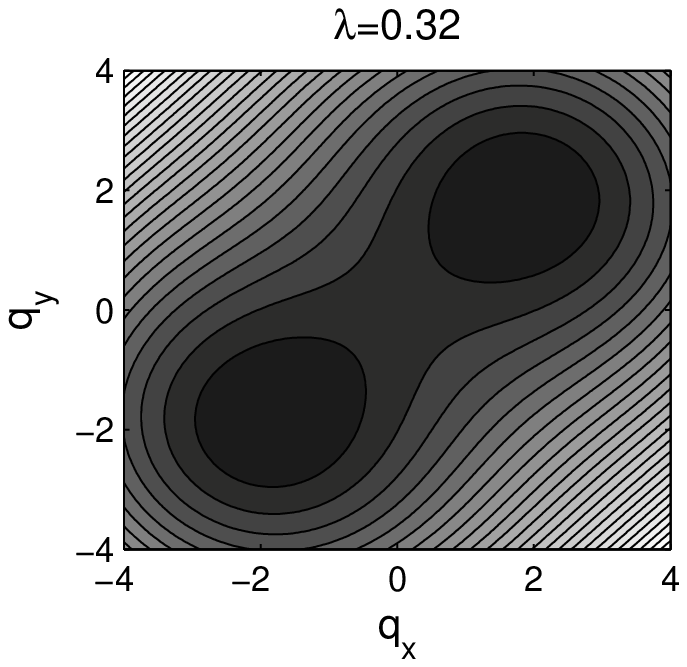} \\
(a) & (b) 
\end{tabular}
\caption{{\bf Fixed point bifurcation of the classical system.} Phase space cross sections in the $q_x - q_y$ plane are shown for the parameter values $\omega=\pi/60,\,\Delta=2\,\text{acot}(2)$,  giving 
the first bifurcation coupling as $\lambda_{b1}=0.26$ and the second bifurcation at the origin when $\lambda_{b2}=0.46$. (a) $\lambda=0.15<\lambda_{b1}$: There is a single stable fixed point located at the origin. Here the fixed point structure is invariant under any reflection axis which passes through the origin. (b)  $\lambda_{b1}<\lambda=0.32<\lambda_{b2}$: Three fixed points occur with a saddle point at the origin and two other fixed points which are stable. There are only two reflection axes passing through the origin which leave the fixed points invariant. At $\lambda=\lambda_{b2}=0.46$ the fixed point at the origin changes from a saddle point to a local maximum, and two new saddle points emerge.} 
\label{levelsfig}
\end{figure}

Our first step is to solve for fixed points of the map by calculating solutions to 
$M(x^{*})=x^{*}$.
The fixed points correspond to periodic orbits of the system in phase space.  
There are trivial fixed points at the equilibrium of the harmonic oscillator where the pseudo-spin is either aligned or anti-aligned with the $z$-axis. These fixed points are the only ones present when $\lambda=0$, one of which is stable,
\begin{align}
s_z=-1/2,\,\,s_x=s_y=
q_x=q_y=p_x=p_y=0.
\label{fp}
\end{align}
while the other ($s_z=1/2$, all other co-ordinates equal to zero) is unstable. As the kicking coupling $\lambda$ is increased a bifurcation of the stable fixed point occurs when  
\begin{eqnarray}
\lambda_b^2=\frac{8\tan(\omega/2)}{\cot(\Delta/2)\pm 1}
\label{bif}
\end{eqnarray}  
where $\omega=\widetilde{\omega}\tau, \Delta=\widetilde{\Delta}\tau, $ are dimensionless variables. Depending on the values of $\omega,\,\Delta$, the above equation can admit either 0,1 or 2 solutions. The smallest value of $\lambda$ for which a solution exists corresponds to a pitchfork bifurcation where the fixed point (\ref{fp}) becomes a saddle point and two new stable fixed points emerge, as illustrated in Fig. \ref{levelsfig}. When there is a second solution to (\ref{bif}) a second pichfork bifurcation occurs at the origin, with the saddle point becoming a local maximum and two new saddle points emerging.

For $\lambda=0$ we can associate the stable fixed point in phase space with the ground state of the quantum system. As $\lambda$ is varied we can (numerically) track this eigenstate of the Floquet operator (\ref{floquet}), which we will refer to as the Pseudo Ground State (PGS), denoted $|\psi_g\rangle$,. Analogously we define the first Pseudo Excited State (PES), denoted $|\psi_e\rangle$.   
Below we demonstrate how the bifurcations in the phase space signal   
distinctive qualitative changes in the PGS, providing insights into a one-body analogue of quantum phase transitions which occur in many-body systems. 

For the oscillator space we choose the basis states which are simultaneous eigenstates of the number operator $N=p^2_x+p^2_y +q^2_x +q^2_y$ and the angular momentum operator $L_z=q_xp_y-q_y p_x$, which we label as $|N,l\rangle, \, N\geq l \geq -N$. The basis $|+\rangle,\,|-\rangle $ for the internal two-level system are eigenstates of $s_z$ with eigenvalues $\pm 1/2$. The unitary map (\ref{floquet}) is invariant under the parity transformation $\Pi= -i\exp(i\pi J_z)$ where $J_z=L_z+s_z$ is the total angular momentum operator. Since $\Pi^2=I$ the Hilbert space splits into  
two parity classes given by
\begin{align*}
O&=\{ |2k,l\rangle|-\rangle,|2k+1,l\rangle|+\rangle \,:\,k\in\mathbb{N}\},\\
E&=\{ |2k,l\rangle|+\rangle,|2k+1,l\rangle|-\rangle \,:\,k\in\mathbb{N}\}.
\end{align*}
As the state space of the system is infinite-dimensional, numerical diagonalisation requires a choice of truncation $N_t$. We take $N_t=18$ giving the dimension of the Hilbert space to be 342, which is expected to give reliable results in the weak coupling limit since for $\lambda=0$ the ground state is $|0,0\rangle |-\rangle\in\, O$. As $\lambda$ is incrementally increased, we numerically solve for the PGS (in the subspace $O$) with an adaptive procedure which requires that the increment $\delta\lambda$ is chosen such that 
$$\left|\langle \psi_g(\lambda)|\psi_g(\lambda+\delta\lambda)\rangle\right|       < 0.01.$$

%\subsection {Husimi functions} 
Husimi functions \cite{Husimi} allow for the representation of quantum states in Hilbert space as a density in a classical phase space, thus providing means for comparison between classical and quantum systems. 
They are defined in terms of coherent states
\begin{equation}
|\alpha_x,\alpha_y\rangle=e^{-|\alpha_x|^2/2}e^{-|\alpha_y|^2/2}\sum_{n_x+n_y=0}^{} \frac{\alpha_x^{n_x}\alpha_y^{n_y}}{\sqrt{n_x!n_y!}}|n_x\rangle|n_y\rangle
\label{cs}
\end{equation}
where $\alpha_\chi=(q_\chi+ip_\chi)/\sqrt{2}$ and $|n_\chi\rangle$ are number eigenstates of 
$N_\chi=p^2_\chi+q^2_\chi$. 
The explicit form of the Husimi function is
\begin{equation*}
\text{H}(\psi,\alpha_x,\alpha_y)=\text{Tr}(\text{Tr}_s(|\psi\rangle\langle\psi|)|\alpha_x,\alpha_y\rangle\langle
\alpha_x,\alpha_y|)
\end{equation*}
which exists in the four-dimensional oscillator phase space.
Above, $\text{Tr}_s$ is the trace over the two-level subsystem.
 \par
\begin{figure}
\begin{tabular}{cc}
\includegraphics[scale=0.55]{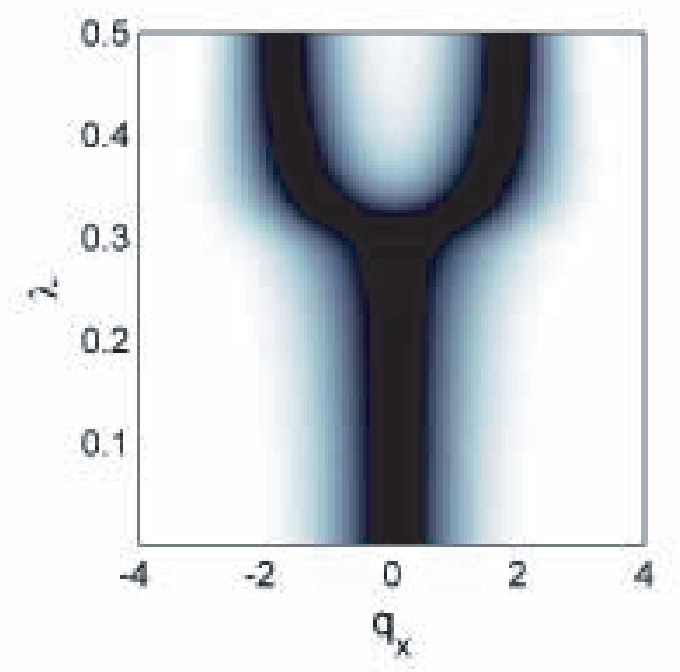} &
\includegraphics[scale=0.55]{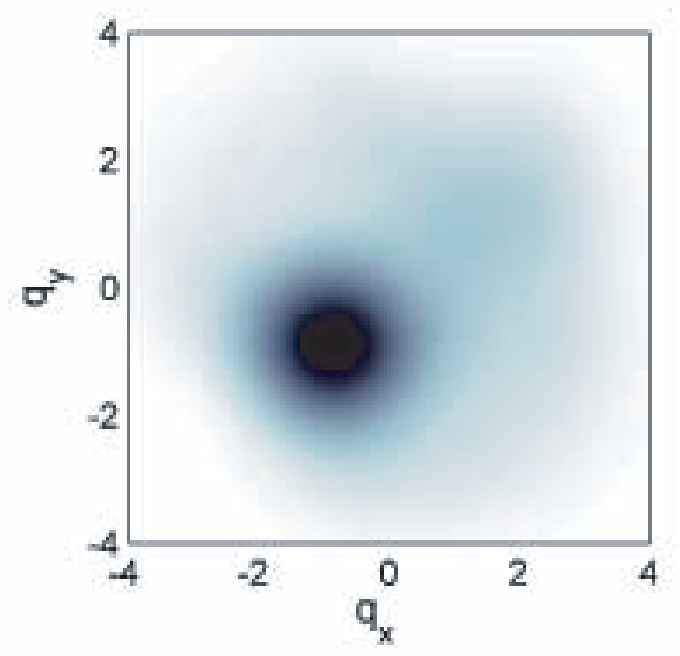} \\
(a) & (b) 
\end{tabular}
\caption{{\bf Husimi function cross sections.} $\omega=\pi/60$ and $\Delta=2\,\text{acot}\,(2)$: (a) The one-dimensional cross section of the Husimi function is shown along the section $q_x=q_y,\,p_\chi=-\tan(\omega/2)q_\chi$ as the coupling $\lambda$ is increased. For small values of $\lambda$ the Husimi function is localised near $q_x=0$. For $\lambda>\lambda_b=0.26$ the Husimi function becomes first delocalise and again localised in two regions of the $q_x$-axis, indicating the wavefunction is a linear combination of localised states which we associate with the fixed points in phase space.  (b) Cross section of the Husimi function in the $q_x - q_y$ plane is shown for the state $| \psi_+\rangle=(|\psi_g\rangle+|\psi_e\rangle)/\sqrt{2}$ when $\lambda=0.32$. The Husimi function is localised at co-ordinates close to the value of the fixed point in phase space, supporting the approximations (\ref{approx1},\ref{approx2}).      }
\label{husimifig}
\end{figure}

  In Fig. \ref{husimifig} the position space cross section of the Husimi function for the PGS is shown as a function of the coupling $\lambda$. A signature of the pitchfork bifurcation of the trivial fixed point is apparent. A state that is highly localised at the origin of the oscillator space splits into branches associated with each of the classical stable fixed points after the bifurcation.
For $\lambda< \lambda_b$ we expect the wavefunction of the PGS to be localised at the stable fixed point in the classical picture; i.e. the PGS is approximately
\begin{equation*}
|\psi_g\rangle\approx |0,0\rangle|-\rangle
\end{equation*}
where $|0,0\rangle$ corresponds to the harmonic oscillator ground state and $|-\rangle$ is the eigenstate of $s_z$ with eigenvalue $-1/2$. After the first bifurcation we expect the PGS to be in a linear combination of two states localised at each of the two new stable fixed points in the classical picture.  A parity invariance of the Floquet operator, determines the phase relationship between these two states with the result that the PGS is approximately
\begin{equation}
|\psi_g\rangle\approx\frac{1}{\sqrt{2}}\left(|\alpha_x,\alpha_y\rangle|\hat{n}\rangle-|-\alpha_x,-\alpha_y\rangle|\hat{n}'\rangle
\right)
\label{approx1}
\end{equation}
and the first PES is approximately 
\begin{equation}
|\psi_e\rangle\approx \frac{1}{\sqrt{2}}\left(|\alpha_x,\alpha_y\rangle|\hat{n}\rangle+|-\alpha_x,-\alpha_y\rangle|\hat{n}'\rangle
\right).
\label{approx2}
\end{equation}
The state $|\hat{n}\rangle= \cos(\theta/2)|+\rangle+e^{i\phi}\sin(\theta/2)|-\rangle$ is determined by the pseudo-spin component
$\hat{n}= \sin(\theta)\cos(\phi){\mathbf i} + \sin(\theta)\sin(\phi){\mathbf j} + \cos(\theta){\mathbf k}$ of the classical fixed point, and $\hat{n}'$ is the parity transformed pseudo-spin component obtained from $\phi \rightarrow \phi+\pi $.

By taking even and odd combinations of the states (\ref{approx1},\ref{approx2}),  we can test this approximation for $\lambda>\lambda_b$. It is expected the two combinations give states localised near each of the classical fixed points respectively, (although with some discrepancy due to mixing of the oscillator reduced density matrix which is absent in the approximation Eq.(\ref{approx2}). The Husimi function of the even combination $(|\psi_g\rangle+|\psi_e\rangle)/\sqrt{2}$ of the numerically determined PGS and first PES is shown in Fig. \ref{husimifig}(b), confirming that localisation occurs.

\begin{figure}
\label{entanglementfig}
\includegraphics[scale=0.6]{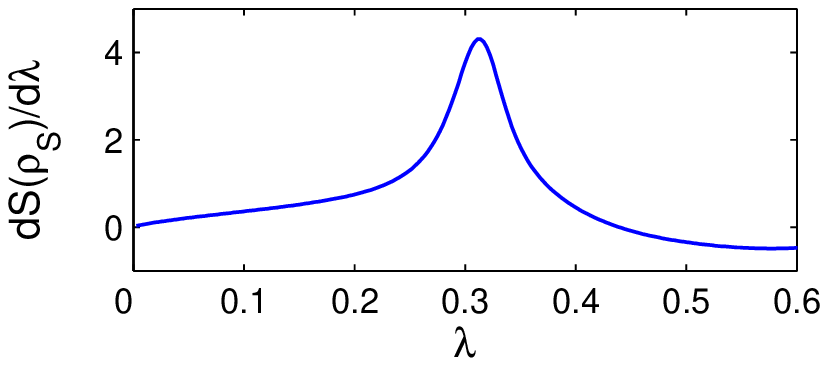}
\includegraphics[scale=0.6]{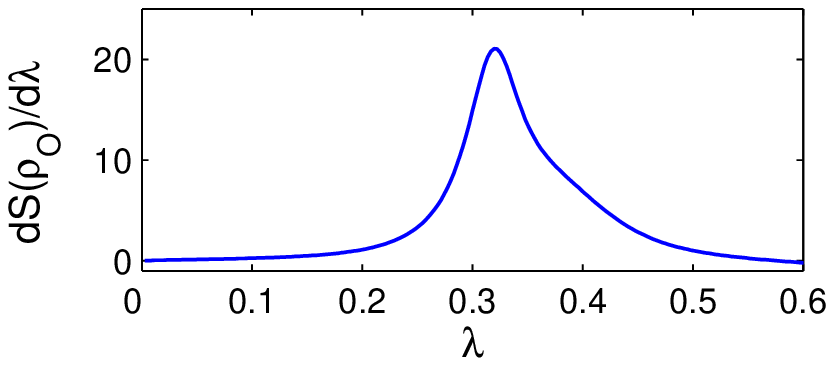}
\includegraphics[scale=0.6]{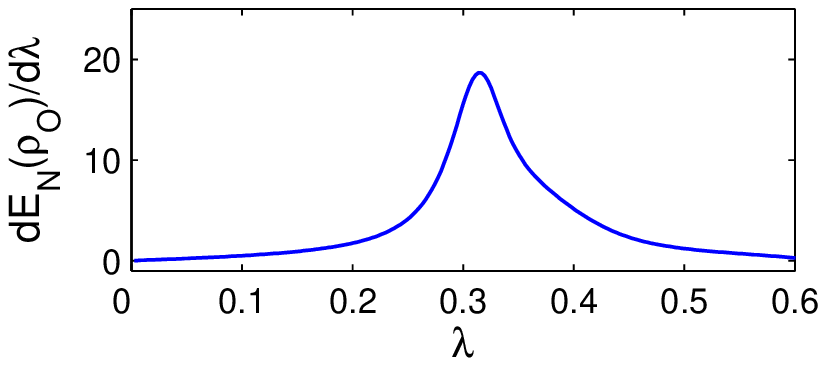}
\caption{{\bf Entanglement measures.} $\omega=\pi/60$, and $\Delta=2\,\text{acot}(2)$: The first derivatives of three entanglement measures for the PGS are shown as a function of $\lambda$. From top to bottom the measures are the von Neumann entropy between the two-level and oscillator subsystems, the von Neumann entropy between the subsystem of a single oscillator degree of freedom and a pseudo-spin/oscillator pair subsystem, and the logarithmic negativity of the mixed-state reduced density matrix of the oscillator degrees of freedom. In all cases the derivative of the entanglement measure has dominant support over the quantum crossover regime $\lambda_{b1}<\lambda<\lambda_{b2}$.}
\end{figure}

As the PGS is a pure state the entanglement between two subsystems $A$ and $B$ can be quantified by the von Neumann entropy of the reduced density matrix 
\begin{equation*}
S(\rho_A)=-Tr(\rho_A\log(\rho_A))\\
\end{equation*}
where $\rho_A \equiv Tr_A(\rho)$.
In this manner the entanglement between the two-level system and the oscillator degrees of freedom, or between an oscillator degree of freedom and the rest of the system, can be calculated from the numerically determined PGS.
To calculate the entanglement between the two oscillator degrees of freedom we use the reduced density matrix  
$
\rho_O=Tr_s(|\psi\rangle\langle\psi|)$.
which is generally a mixed state. In this instance the entanglement may be quantified through the logarithmic negativity \cite{Vidal2002}
\begin{equation*}
E_{{N}}(\rho_O)\equiv \log_N\parallel\rho^{T_A}\parallel_1
\end{equation*}
where the norm is the trace norm $\parallel A\parallel_1=Tr\sqrt{A^{\dagger}A}$ and $\rho^{T_A}$ denotes the partial transpose over subsystem $A$.

The model of this paper might be realised in a cylindrical Paul trap. In order to ensure that the motion in the $x-y$ plane is well confined we need to ensure that the secular frequencies in this plane $\omega_{x,y}$ are  smaller than the frequency in the $z-$direction.  Using the stability diagram for this kind of trap (see figure 1 in \cite{Leibfried}) we can operate at trap parameters such that $\omega_{x,y}/\omega_z \approx 10$ without significant micromotion in the $x-y$ motion.  Although this requires operation close to the stability edge it should be achievable. In order to couple different components of the Pauli matrix to the vibrational degree of freedom we can make use of the Raman scheme for state dependent displacements introduced by Monroe et al.\cite{Monroe}. Consider first that the Raman beams are directed along the $x-$axis of the trap but the same arguments hold for the $y-$ axis Raman pulses. The effective interaction Hamiltonian for the Raman pulses is  $H_R=\chi q_x\sigma_z$ where $\sigma_z=|e\rangle \langle e|-|g\rangle\langle g|$ with $|e\rangle,|g\rangle$ are the ground and excited states of the relevant two-level electronic transition. In other words we have a pseudo-spin realisation with $|+\rangle=|e\rangle,\ |-\rangle=|g\rangle$.   If we `sandwich' this Raman pulse by two laser pulses tuned to the carrier transition with an appropriate phase choice, the displacement can be made to depend on any component of the Pauli matrices. For example the pulse sequence,
\begin{equation}
e^{-i\pi/4\sigma_y} e^{-i\chi q_x\sigma_z} e^{i\pi/4\sigma_y}=e^{-i\chi q_x\sigma_x}
\end{equation}
will achieve the desired result for the $x-$ kick. 

We are now in a position to consider the experimental signatures of our results. As a specific example, we will focus on detecting the entanglement implicit in the state just beyond the first bifurcation, approximately given by in Eq.(\ref{approx2}). This can be reached adiabatically starting in the state in $|0,0,\rangle|-\rangle$and using a sequence of pulses with gradually increasing coupling strength. At the end of a sequence the probability to detect the atom in the excited state is simply
\begin{equation}
P(+)=\cos^2(\theta/2)\left (1- e^{-2\alpha_x^2-2\alpha_y^2}\right )
\end{equation}
where $\theta, \alpha_x, \alpha_y$ are determined by the classical fixed points.  Sampling this distribution thus indicates the support on the classical fixed points. Readout of the excited state is easily done in ion trap realisations using a cycling transition\cite{Leibfried}. The experimental verification of the entanglement that results after the bifurcation is therefore quite achievable with current technology.

\acknowledgements
This work was supported by the Australian Research Council and the European Union IP QAP.

\end{document}